\documentclass[prl,twocolumn,superscriptaddress,showpacs,amsmath,amssymb,floatfix,nofootinbib]{revtex4-2}
\usepackage{graphicx}
\usepackage{bm}
\usepackage[hidelinks]{hyperref}
\newcommand{\Ar}[1]{$^{#1}$Ar}
\newcommand{\slabtally}{inclusive neutron-removal tally}
\newcommand{\fig}[2]{\IfFileExists{#1}{\includegraphics[width=\columnwidth]{#1}}{\fbox{\parbox[c][#2][c]{0.95\columnwidth}{\centering\small [#1: run \texttt{make\_figs.py} on the VM]}}}}

\begin{document}

\title{Trace isotopes set the neutron horizon of liquid argon}

\author{Yasser Maghrbi}
\email[Contact author: ]{yasser.maghrbi@oca.eu}
\affiliation{Universit\'e C\^ote d'Azur, Observatoire de la C\^ote d'Azur, ARTEMIS, CNRS, Nice, France}
\affiliation{Gulf University for Science and Technology, Kuwait}

\begin{abstract}
Near 57 keV, destructive interference makes $^{40}$Ar almost transparent to neutrons. The 0.4\% of $^{36}$Ar and $^{38}$Ar in natural argon then supply over 90\% of its opacity, amplified 230-fold. Published transmission data rule out the $^{36}$Ar evaluation FLUKA uses by default, even with every other isotope made transparent, and bound its cross section in the window to 2--11~b. In a DUNE-scale detector, switching to the data-consistent evaluations moves the fiducial fraction of $^{41}$Ar calibration captures by 9--13\% while the total yield changes by only 2\%: the effect is on where captures occur, not how many.
\end{abstract}

\maketitle

A neutron captured on \Ar{40} releases a cascade of $\gamma$ rays carrying about 6.1~MeV in total~\cite{aced}. For the Deep Underground Neutrino Experiment (DUNE), whose far detector will hold four 17-kt liquid-argon modules~\cite{dune_tdr2,dune_vd}, this final state has two opposite meanings. It is the signal of the pulsed-neutron calibration planned for the far detector~\cite{dune_tdr4,dune_idr,coldbox,conlon}, and it is a background to low-energy neutrino analyses when the neutron comes from the rock or from cosmic-ray secondaries~\cite{capozzi,zhu,dune_phase2,meighen}. In both cases position matters: a capture near a wall and a capture deep in the fiducial volume are not equivalent. The same transport problem arises in the Fermilab short-baseline program~\cite{sbn}, in DarkSide-20k and DEAP-3600~\cite{ds20k,deap_det}, in the argon shield of LEGEND-1000~\cite{legend}, and in the reconstruction of neutron energy in liquid-argon time-projection chambers~\cite{friedland,castiglioni,microboone_n,blips}. The question is simple and still open: how far can a neutron travel in liquid argon before it is captured?

DUNE's neutron-calibration concept relies on an unusual piece of argon nuclear physics. At most energies the total neutron cross section of \Ar{40} is a few barns (1~b $=10^{-24}$~cm$^2$), giving a mean free path well below a meter. Near 57~keV it collapses to 0.42~mb, because scattering from the nuclear potential interferes destructively with a broad $s$-wave resonance near 76~keV~\cite{winters,liou,mughabghab}. A 2.45-MeV neutron from a deuterium--deuterium generator first scatters and slows down; the small fraction that reaches this transparency window can then fly very far. This is why a source on the cryostat roof can illuminate a much larger volume~\cite{dune_tdr4,dune_idr}. The design literature quotes 859~m for \Ar{40}~\cite{dune_tdr2,dune_idr}, more than 2~km when only elastic scattering is counted~\cite{dune_tdr4}, and about 30~m for natural argon~\cite{dune_tdr2,dune_tdr4}. These numbers do not contradict one another; they reflect different assumptions about what fills the window.

That distinction matters because natural argon is not pure \Ar{40}. It contains 0.334\% \Ar{36} and 0.063\% \Ar{38}~\cite{iupac}, and neither isotope shares the deep minimum at 57~keV. Their cross sections there are of order barns, and they are poorly known. The only transmission measurement on enriched \Ar{36} stops at 6~keV~\cite{chrien}, and neutron capture on \Ar{36} and \Ar{38} above thermal energy was first measured in 2018~\cite{tessler}. Evaluated libraries therefore differ by more than an order of magnitude for \Ar{36} in the window~\cite{jeff33,endf81,jendl5,jeff40}. With the JEFF-3.3 files that FLUKA loads by default~\cite{fluka_battistoni,fluka_ahdida}, the long-flight scale of natural argon is 16~m; with ENDF/B-VIII.1 it is 42~m. The issue is not confined to one code. Geant4~\cite{allison} removed its \Ar{36} and \Ar{38} files in 2024 after finding that they inflated the argon cross section by up to 50\%~\cite{g4_notes}; those files were converted from JEFF-3.3~\cite{g4_data}, the evaluation FLUKA still loads by default, and a 2026 user report traced neighboring-isotope substitutions that appear when those files are absent~\cite{neuberger}.

At the minimum, the usual abundance hierarchy reverses: the isotope that makes up 99.6\% of the liquid almost steps out of the transport problem, exposing the other 0.4\%. We use this to isolate the trace-isotope data, to connect the nuclear minimum with the long-distance transport scale, and to ask what survives in a finite detector of DUNE size.

\begin{figure}[t]
\fig{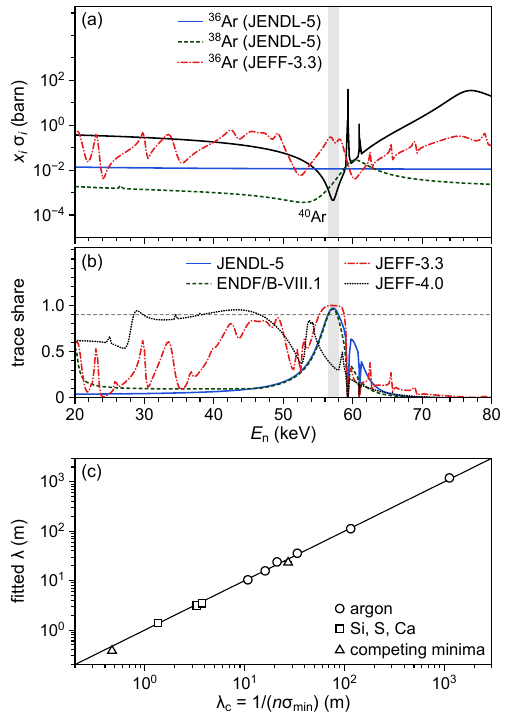}{12.2cm}
\caption{The host minimum exposes the trace isotopes. (a)~Abundance-weighted total cross sections of \Ar{40}, \Ar{36} and \Ar{38} (JENDL-5), with the \Ar{36} term of the JEFF-3.3 file used by default in FLUKA. Near 57~keV the \Ar{40} term falls more than an order of magnitude below the trace terms. (b)~Share of the natural-argon cross section supplied by \Ar{36} and \Ar{38}. Cross sections are averaged over 0.25-keV intervals; the narrow \Ar{40} resonances at 59.3 and 60.9~keV are shown pointwise. The shaded band marks where it exceeds 90\% in JENDL-5; JEFF-4.0 has its own \Ar{40} file with a displaced minimum. (c)~Neutron horizon fitted from FLUKA absorption tails against Eq.~(\ref{eq:rule}), for the media of Table~\ref{tab:rule}; the line is equality.}
\label{fig:mechanism}
\end{figure}

For natural argon, $\sigma_{\rm nat}=0.996\,\sigma_{40}+0.0034\,\sigma_{36}+0.0006\,\sigma_{38}$. The \Ar{40} file shared by JEFF-3.3, ENDF/B-VIII.1, JENDL-5 and BROND-3.1~\cite{jeff33,endf81,jendl5,brond} reaches 0.42~mb at 57.15~keV, of which 0.19~mb is radiative capture. In the evaluations consistent with the transmission data discussed below, the abundance-weighted trace terms add about 11--13~mb there [Fig.~\ref{fig:mechanism}(a)]. One atom in 250 therefore supplies more than 90\% of the opacity around the minimum [Fig.~\ref{fig:mechanism}(b)], and relative to its abundance the trace component is amplified about 230-fold. We call this antiresonance amplification: a minimum of the host cross section turns a measurement on the natural material into a probe of isotopes that are otherwise too dilute to see. The same mechanism appears in the 24- and 54-keV neutron-filter windows of iron and silicon, where minor isotopes partly fill the host minimum~\cite{mughabghab,block_brugger,annri}. Argon is especially clean because its minimum is so deep.

The same minimum sets the long-distance transport scale. We define the neutron horizon $\lambda_c$ as the exponential decay length of the far tail of the absorption distribution around a point source in an effectively unbounded medium; it is neither the median capture distance nor the reach of neutrons in a particular detector. To survive to a large distance $R$, a neutron must make its long flight near the energy where the total cross section is smallest, since elsewhere the factor $e^{-n\sigma(E)R}$ suppresses it. At large $R$ the smallest cross section therefore fixes the exponential slope, the standard deep-penetration result~\cite{goldstein}:
\begin{equation}
\lambda_c = \frac{1}{n\,\sigma_{\min}},
\label{eq:rule}
\end{equation}
where $n=2.10\times10^{22}$~cm$^{-3}$ is the atomic density of liquid argon and $\sigma_{\min}$ the lowest total cross section below the source energy. Nothing is fitted in Eq.~(\ref{eq:rule}). FLUKA simulations of 2.45-MeV neutrons in volumes large enough to contain the tail agree with it to 2--13\% for single-minimum argon cases from 16~m to 1.2~km, and to 1--9\% for silicon, sulfur and calcium [Fig.~\ref{fig:mechanism}(c) and End Matter]. We do not mean Eq.~(\ref{eq:rule}) as a universal transport law. Once a long-flight minimum exists, however, its depth fixes the far-tail slope; in natural argon that depth is set mainly by \Ar{36} and \Ar{38}.

\begin{figure}[t]
\fig{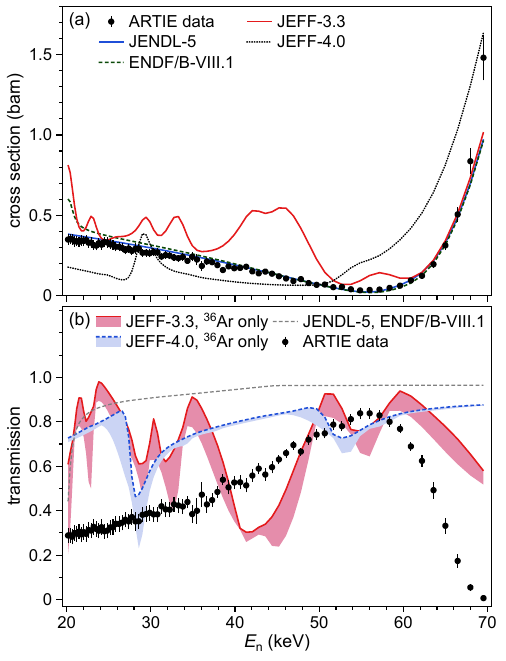}{6.1cm}
\caption{ARTIE data against the evaluations. (a)~Cross sections from the ARTIE transmission~\cite{artie,exfor14824} with statistical errors, and the four evaluations folded through a 300-ns time-of-flight response (no nuisance adjustment). (b)~A transparent-host test of the default \Ar{36} evaluation. Points: ARTIE transmission, conservatively at the lower edge of the total uncertainty, with statistical errors. Bands: the largest transmission that the \Ar{36} file alone allows when \Ar{40} and \Ar{38} are made perfectly transparent, for JEFF-3.3 (solid outline) and JEFF-4.0 (dashed outline), over the response family of the End Matter. Where the data lie above a band, no choice of \Ar{40} can rescue that file. The JENDL-5 and ENDF/B-VIII.1 ceilings (dashed) lie above every point.}
\label{fig:ceiling}
\end{figure}

The ARTIE experiment measured neutron transmission through 1.68~m of liquid argon between 20 and 70~keV~\cite{artie}; its 76 points are in EXFOR~\cite{exfor14824}. ARTIE compared the data with one evaluation and confirmed the dip; which isotopes fill it was not part of that analysis. Because the target is thick and the energy response matters, we compare an evaluation with the data only after folding the transmission $e^{-N\sigma(E)}$, where $N$ is the target areal density, through the time-of-flight response and profiling the quoted density, normalization and energy-scale uncertainties (End Matter). For the 39 points between 30 and 60~keV, JENDL-5 and ENDF/B-VIII.1 give $\chi^2=42$ and 59, JEFF-3.3 and JEFF-4.0 give 1656 and 3073 [Fig.~\ref{fig:ceiling}(a)]. With normalization and offset left free, the values are 16, 16, 407 and 863. The problem is spectral shape, not normalization.

The fold still depends on the \Ar{40} model. A separate positivity bound removes that dependence. Since cross sections cannot be negative, natural argon cannot transmit more neutrons than its \Ar{36} alone would if every other isotope were perfectly transparent. We therefore set the \Ar{40} and \Ar{38} opacity to zero and compute the largest transmission the JEFF-3.3 \Ar{36} file can allow [Fig.~\ref{fig:ceiling}(b)]. At 45~keV this ceiling lies between 0.24 and 0.42 across the response family, whereas ARTIE measured 0.67 (0.63 at the lower edge of its total uncertainty). In 9 of the 12 bins between 40 and 50~keV the data sit above the ceiling for every response in the family, and in all 12 with the responses that reproduce the measured spectrum (End Matter); a second, weaker violation appears at 54--56~keV. Put plainly, the \Ar{36} file that FLUKA uses by default would on its own have removed more neutrons from the beam than the entire argon sample did. This exclusion does not require the \Ar{40} cross section to be right.

Independent data tell the same story at other energies (End Matter, Table~\ref{tab:anchors}). The free-atom thermal scattering cross section of \Ar{36} is 73.7~b~\cite{sears}, against 220.5~b in JEFF-3.3 and 141.0~b in JEFF-4.0, while ENDF/B-VIII.1 and JENDL-5 give 73.6 and 70.3~b. The natural-argon transmission data of Enik \textit{et al.}~\cite{enik} also reject the large JEFF-3.3 structure below 6~keV. Their point at 30.1~keV is more revealing: JEFF-4.0 and JENDL-5 both reproduce the measured total, $0.281\pm0.021$~b, yet attribute about 88\% and 4\% of it to \Ar{36}. Agreement in a bulk total can hide a wrong isotopic decomposition. The \Ar{40} minimum removes that accidental compensation and exposes the trace component.

The ARTIE spectrum also constrains the size of the trace opacity. Letting the \Ar{40} contribution move by 10--20\%, scanning the response model, and fitting a smooth correction to the JENDL-5 \Ar{36} shape gives an effective \Ar{36} cross section of 2.4--10.7~b at 45--55~keV and at most about 11~b at 55--65~keV, with \Ar{38} held fixed~\cite{sm}. JENDL-5 and ENDF/B-VIII.1 lie inside this range; the JEFF evaluations do not. The corresponding horizon of atmospheric argon is 12--50~m in this fit and 34--42~m in the two evaluations that reproduce the window. The 16-m value of the FLUKA default comes from a \Ar{36} file that the transmission data exclude; the number itself is not forbidden for a different spectral shape.

\begin{figure}[t]
\fig{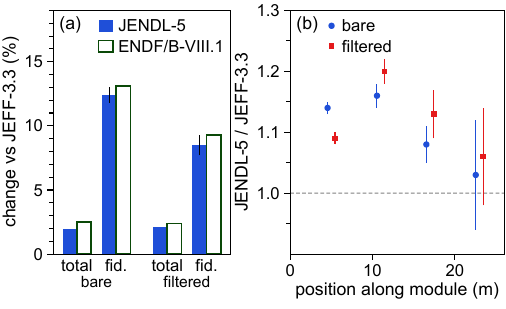}{4.6cm}
\caption{Where, not how many. (a)~Change relative to JEFF-3.3 in the total \Ar{41} yield and in the fiducial acceptance of \Ar{41} captures (the fraction more than 2~m from every wall), for a bare and a filtered source on the roof of a DUNE-scale module. (b)~Ratio JENDL-5/JEFF-3.3 of the \slabtally{} in 6-m slabs along the first 26~m of the module (the statistically limited last slab is given in the Supplemental Material); the source port is at 3~m. Errors are cycle-to-cycle standard errors.}
\label{fig:detector}
\end{figure}

\textit{Where, not how many.}---A material length is not yet a detector observable. We modeled a DUNE-scale volume of liquid argon, $14.5\times12\times58$~m$^3$, inside the cryostat layers, with a 2.45-MeV deuterium--deuterium source on a sealed roof port as in the calibration design~\cite{dune_tdr4,conlon}, both bare and inside the proposed sulfur--iron--lithium filter. The design requires the source to reach most of the detector and verifies this by simulation~\cite{dune_tdr4}. To remove the uncertain source intensity, we use a fiducial acceptance: the fraction of $^{40}$Ar(n,$\gamma$)$^{41}$Ar calibration captures that occur more than 2~m from every wall.

Replacing JEFF-3.3 by JENDL-5 changes the total \Ar{41} yield by only 2.0\%, but changes this acceptance by $12.4\pm0.6$\% for the bare source and $8.5\pm0.8$\% for the filtered one; ENDF/B-VIII.1 gives 13.1\% and 9.3\% [Fig.~\ref{fig:detector}(a)]. The inclusive neutron-removal tally shows the same spatial pattern, changing by roughly 8--20\% over the first 20~m [Fig.~\ref{fig:detector}(b)]. The change is largest within about 15~m of the source and fades toward the far end: a module only 12--15~m across, smaller than either horizon, does not see the far tail itself, but it does see where the captures move. A simulation tuned only to the total capture rate can therefore look right while its calibration map is wrong.

This is directly testable within the calibration program. The DUNE Vertical Drift ColdBox pulsed-neutron-source study already compares neutron-induced data with FLUKA-based simulation~\cite{coldbox}. A fiducial fraction, or a lower-to-upper capture ratio, adds the advantage that the generator strength cancels. In our ProtoDUNE-sized model~\cite{protodune_perf,bordoni} the lower-to-upper ratio beneath a roof source changes by 15\% between JEFF-3.3 and JENDL-5~\cite{sm}, a source-normalization-free target for a future generator run. The same saturation is visible with underground argon: depleting \Ar{36} as in argon from deep wells~\cite{xu_uar}, and \Ar{38} by the same factor, moves the infinite-medium horizon from 34~m to about 0.7~km, yet changes the fiducial acceptance of the DUNE-scale module by only about 8\%, because the detector is already smaller than the material transport scale~\cite{sm}.

Two measurements would sharply reduce the remaining uncertainty. Transmission through argon enriched in \Ar{36}, and separately in \Ar{38}, between about 1 and 100~keV would break the isotopic degeneracy of natural argon, and measurements at more than one target thickness would separate unresolved resonance structure from a genuine opacity floor~\cite{froehner}. The MArEX program at CERN~\cite{marex} and the ARTIE-II target described by Erjavec~\cite{erjavec} are natural routes. On the detector side, the spatial ratios above need no absolute source calibration and can be measured in the same pulsed-source program.

For present simulations, the practical lesson is that the choice of argon evaluation is not an invisible implementation detail. The spread between ARTIE-consistent evaluations should be propagated through spatial calibration observables, and the default JEFF-3.3 \Ar{36} file should not be regarded as validated in this energy range. The geometries, isotopic compositions, processed-file identifiers, checksums and \Ar{41} yields form a benchmark that can be reproduced in FLUKA, Geant4 or LArSoft~\cite{sm,iaea_g4} and are available from the author; this turns the nuclear-data question into a detector test rather than a code comparison.

The mechanism reaches beyond argon. Interference minima in iron and silicon are likewise filled by minor isotopes; in the evaluated files, enrichment lengthens their long-flight scales by more than an order of magnitude, and FLUKA transport follows the same trend~\cite{sm}. Antiresonance amplification thus offers a general way to expose isotope data that stay hidden in measurements of a natural element away from the host minimum.

Near 57~keV, \Ar{40} steps aside. The one argon atom in 250 that is not \Ar{40} then sets the scale of the longest neutron flights. In a finite detector that microscopic fact shows up not so much in how many calibration captures occur as in where they occur. The interference window that hides the host thus exposes both the trace-isotope nuclear data and a detector systematic that can be measured in situ.

\section*{End Matter}

\textit{Transport and the horizon rule}---All simulations used FLUKA 4-5.2~\cite{fluka_battistoni,fluka_ahdida} with pointwise neutron transport below 20~MeV and neutrons followed to thermal energy; captures were scored as residual nuclei in concentric shells or rectangular regions; the tail fits use an inclusive terminal-reaction tally (\Ar{37}, \Ar{39}, \Ar{41} and the sulfur and chlorine residuals), the calibration results \Ar{41} alone. A library other than the default was used by replacing the argon files of the JEFF-3.3 directory with those of JENDL-5, ENDF/B-VIII.1 or JEFF-4.0 and checking the replacement byte by byte; processed files were checked against the original evaluations with JANIS~\cite{janis}, and experimental data were taken from EXFOR~\cite{exfor}. Isotopic compositions were set with material and compound cards. A 2.45-MeV point source sat at the center of argon spheres or boxes large enough to contain essentially the whole fitted tail (150~m for atmospheric argon, 1.2~km for underground argon, 15~km for pure \Ar{40}). The outer terminal-reaction distribution was fitted with a pure exponential and with an exponential divided by $\sqrt{R}$, the form expected near a smooth minimum; the better fit is reported. Each case ran in 5--10 independent cycles of $1$--$2\times10^6$ neutrons, and errors are cycle-to-cycle standard errors. Table~\ref{tab:rule} compares the fitted lengths with Eq.~(\ref{eq:rule}). Where several minima of similar depth compete (iron, and argon with the JEFF-4.0 files), finite-range tails fall below the single-minimum prediction.

\begin{table}[h]
\caption{The rule $\lambda_c=1/(n\sigma_{\min})$ against FLUKA. Lengths in meters.}
\label{tab:rule}
\begin{ruledtabular}
\begin{tabular}{llcc}
Medium & Data & Predicted & Fitted \\
\hline
atmospheric Ar & JEFF-3.3 & 16.1 & 15.7--15.9 \\
atmospheric Ar & JENDL-5 & 33.8 & 35.9 \\
Ar, \Ar{36} at 0.1\% & JEFF-3.3 & 21.2 & 24.0 \\
Ar, \Ar{36} at 1\% & JEFF-3.3 & 10.8 & 10.4 \\
underground Ar & JEFF-3.3 & 116 & 112--114 \\
pure \Ar{40} & shared file & 1132 & 1195--1210 \\
atmospheric Ar & JEFF-4.0 & 25--30\footnotemark[1] & 24 \\
natural Si & JEFF-3.3 & 1.36 & 1.34--1.43 \\
natural S & JEFF-3.3 & 3.32 & 3.04--3.25 \\
natural Ca & JEFF-3.3 & 3.75 & 3.40--3.61 \\
natural Fe & JEFF-3.3 & 0.47 & 0.39--0.41\footnotemark[1] \\
\end{tabular}
\end{ruledtabular}
\footnotetext[1]{Several competing minima.}
\end{table}

\textit{Folding the ARTIE data}---The 76 cross sections (20--70~keV) are those of EXFOR entry 14824~\cite{exfor14824}; an independent extraction from the vector graphics of the published figure~\cite{artie} agrees with them to better than 0.2\%. Statistical uncertainties, which the entry does not tabulate separately, were read from the figure; where a statistical bar is hidden by its marker, half the marker height is used as an upper bound. For each bin, the transmission $e^{-N\sigma(E)}$ with $N=3.30$ atoms/b is averaged over the neutron energies that fall into that 200-ns bin at 63.82~m and converted back into an apparent cross section, as for the data. The energy response was modeled by a Gaussian timing spread of 150--600~ns with an optional 1-$\mu$s delayed tail carrying up to 20\% of the weight; nuisance parameters are the areal density (1.3\% prior), an additive offset (0.012~b, the 4\% transmission systematic of the experiment) and the energy scale (0.3\%). Closure tests on synthetic spectra verified the procedure.

\textit{The transparent-host ceiling}---In every bin the measured transmission is an average of $e^{-N(x_{40}\sigma_{40}+x_{36}\sigma_{36}+x_{38}\sigma_{38})}$ over the response. Because $\sigma_{40},\sigma_{38}\ge0$, it cannot exceed the average of $e^{-N x_{36}\sigma_{36}}$. We evaluated this bound with $N=3.26$ atoms/b and $x_{36}$ 1\% below nominal, using the skew-normal energy response fitted in Ref.~\cite{erjavec}: location at the nominal bin energy, scale 2.35~keV and shape 3.11 at 54.75~keV, which give a 16--84\% width of 3.0~keV, consistent with the 2.9~keV tabulated there. The scale was extrapolated as $(E/54.75\,{\rm keV})^p$ with $p=1$--2 and inflated up to twofold, and the bound was compared with the data at the lower edge of the total uncertainty. Centring this response on its mean, or using the Gaussian timing responses of the fits, lowers the ceiling further and excludes the file in all 12 bins. Pulls are per bin and are not combined, since neighboring bins share the response.

\begin{table}[h]
\caption{Evaluations against measurements (measured value in parentheses). Thermal values are free-atom cross sections~\cite{sears}; the \Ar{40} thermal capture, 0.66~b in all files, agrees with Refs.~\cite{sears,aced}.}
\label{tab:anchors}
\footnotesize\setlength{\tabcolsep}{1.2pt}
\begin{ruledtabular}
\begin{tabular}{lcccc}
Test & JEFF-3.3 & JEFF-4.0 & ENDF & JENDL \\
\hline
ARTIE $\chi^2$ (39 pts) & 1656 & 3073 & 59 & 42 \\
\quad free norm./offset & 407 & 863 & 16 & 16 \\
\Ar{36}-only ceiling & fails & weak & passes & passes \\
\Ar{36} thermal (73.7 b) & 220.5 & 141.0 & 73.6 & 70.3 \\
Ar 1.5--5.9 keV (0.52--0.57 b) & 2.0--2.6 & 0.4--0.7 & 0.6--0.7 & 0.6--0.7 \\
Ar 30.1 keV (0.281 b) & 0.44 & 0.28 & 0.31 & 0.30 \\
\Ar{36} MACS (1.9 mb) & 8.5 & 11.9 & 8.9 & 5.2 \\
\Ar{40} th.\ elastic (0.40 b) & 0.647 & 0.350 & 0.647 & 0.647 \\
\end{tabular}
\end{ruledtabular}
\end{table}

\textit{Detector models}---The module is $14.5\times12\times58$~m$^3$ of liquid argon under 0.8~m of argon gas, enclosed by a 1.2-mm stainless-steel membrane, 80~cm of foam insulation modeled as CH$_2$ at 0.09~g\,cm$^{-3}$ (14\% hydrogen by mass, an upper bound for polyurethane) and 12~mm of carbon steel, with a 30-cm-radius port 3~m from one end closed by a 2-cm flange; the isotropic 2.45-MeV source sits 5~cm above the flange, bare or inside 1~cm of $^6$Li, 20~cm of sulfur and 18~cm of iron with a 10-cm nickel reflector~\cite{dune_tdr4,conlon}. The fiducial region lies more than 2~m from every wall and is divided into nine 6-m slabs and three depth bands. Fiducial acceptances use \Ar{41} only; Fig.~\ref{fig:detector}(b) shows the \slabtally{}. Details of the underground-argon and ProtoDUNE-sized models are given in the Supplemental Material~\cite{sm}.

\textit{AI disclosure}---Claude Opus 5.5 (Anthropic) and ChatGPT GPT-5.6 Sol (OpenAI) assisted with literature searches and drafting and editing of the text and analysis code. The author conceived the study, designed and ran all simulations, performed the analysis, drew the conclusions, verified every number, script and reference, and takes full responsibility for the content.

\end{document}